\documentclass{raa}           
\usepackage{lineno}
\usepackage{graphicx,times}
\usepackage{natbib}
\usepackage{amssymb,amsmath,chemfig}
\bibpunct{(}{)}{;}{a}{}{,}

\usepackage[pagebackref=true]{hyperref}

\begin{document}

   \title{JWST observations constrain the time evolution of fine structure constants and dark energy-electromagnetic coupling}

 \volnopage{ {\bf 20XX} Vol.\ {\bf X} No. {\bf XX}, 000--000}
   \setcounter{page}{1}

   \author{Ze-Fan Wang
   \inst{1, 2}, Lei Lei\inst{1, 2}, Lei Feng\inst{1, 2}, Yi-Zhong Fan\inst{1, 2}
   }

   \institute{ Key Laboratory of Dark Matter and Space Astronomy, Purple Mountain Observatory, Chinese Academy of Sciences, Nanjing 210033, China; {\it  fenglei@pmo.ac.cn, leilei@pmo.ac.cn}\\
        \and
             School of Astronomy and Space Science, University of Science and Technology of China, Hefei 230026, China\\
\vs \no
   {\small Received 20XX Month Day; accepted 20XX Month Day}
}

\abstract{It was hypothesized in the literature that some physical parameters may be time-evolving and the astrophysical data can serve as a probe. Recently, James Webb Space Telescope (JWST) have released its early observations. In this work, we select the JWST spectroscopic observations of the high redshift ($z>7.1$) galaxies with strong [\ion{O}{III}] ($\lambda=4959$\AA \,and $5007$\AA \,in the rest frame)  emission lines to constraint the evolution of the fine structure constant ($\alpha$). With the spectra from two galaxies at redshifts of $7.19$ and $8.47$, the deviation of $\alpha$ to its fiducial value is found to be as small as $0.44^{+8.4+1.7}_{-8.3-1.7} \times 10^{-4}$ and $-10.0^{+18+1.5}_{-18-1.5} \times 10^{-4}$, respectively (the first error is statistical and the latter is systematic). The combination of our results with the previous data reveals that $\frac{1}{\alpha} \frac{d \alpha}{dt} = 0.30^{+4.5}_{-4.5} \times 10^{-17}~{\rm yr^{-1}}$.  Clearly, there is no evidence for a cosmic evolution of $\alpha$.  The prospect of further constraining the time evolution of $\alpha$ is also discussed. The scalar field of dark energy is hypothesized to drive the acceleration of the universe's expansion through an interaction with the electromagnetic field. By integrating the observational data of the fine-structure constant variation, $\frac{\Delta\alpha}{\alpha}(z)$, we have established a stringent upper limit on the coupling strength between dark energy and electromagnetism. Our analysis yields $\zeta \leq 3.92 \times 10^{-7}$ at the 95\% confidence level, representing the most stringent bound to date.  
}

   \authorrunning{Z.-F. Wang et al. }            
   \titlerunning{JWST constrain $\alpha$ and DE-EM coupling time evolution}  
   \maketitle

%
\section{Introduction}           
\label{sect:intro}

Fundamental physical constants are one facet of nature's laws, but are they really non-revolving in the universe? 
\cite{DIRAC1937} proposed the famous large-number coincidence, suggesting an association between fundamental constants and the current status of the universe. About one decade after that, \cite{teller1948change} argued that the time variation of gravitational constant ($G$) seems impossible due to the ecosystems on our Earth. Nevertheless, further probe is still necessary to check whether there is the cosmic evolution of the physical constants or not. One good target is the fine-structure constant which can be expressed as $\alpha = \frac{e^2}{4 \pi \varepsilon_0 \hbar c}$, where $e$ is the electron charge, $\varepsilon_0$ is the vacuum permittivity, $\hbar$ is the reduced Planck constant, and $c$ is the speed of light in the vacuum. The numerical value of this dimensionless constant is found to be $\alpha^{-1} = 137.035999206(11)$  (\citealt{Morel2020}). A reliable identification of a deviation of $\alpha$ from such a standard value would suggest the presence of the new physics. 

In the middle of the $20^{th}$ century, \cite{stanyukovich1963possible} and \cite{gamow1967electricity} introduced the idea of time-variation of $\alpha$ in cosmology. \cite{dyson1967time} deduced that the time variation of $e$ is less than $1 / 1600$ during the history of Earth from the terrestrial existence of the nuclei $Re^{187}$ and $Os^{187}$. But \cite{gamow1967electricity} did give suggestions on the detection of the time-varying $\alpha$ through astronomical sources.  \cite{Savedoff1956} firstly analyzed the spectral fine-structure of the emission lines of [\ion{N}{II}] and [\ion{Ne}{III}] in the spectrum of the nearby Seyfert galaxy. Astronomers have since been clear that the upper bound of the relative variation $\Delta \alpha / \alpha$ and the time variation $\frac{1}{\alpha} \frac{d\alpha}{dt}$ in our local universe must be extremely small \citep{PhysRevC.74.064610, doi:10.1126/science.1154622, 10.1093/mnras/stv2148, martins2017stability}. Consequently, modern observations focus on high-redshift celestial bodies to explore the potential change of $\alpha$ in the early universe \citep{PhysRevD.97.023522, doi:10.1126/sciadv.aay9672}.
Thanks to the successful launch and outstanding performance of the James Webb Space Telescope (JWST), astronomers are now able to catch a glimpse of the young generation of galaxies and stars at very high redshifts. Recently, \citealt{2024ApJ...968..120J,2024arXiv240508977J}\footnote{After the initial submission of our manuscript for publication on May 4, 2024, Jiang et al. 2024b appeared in arXiv. In our analysis,  just the high resolution spectral data have been taken into account, while Jiang et al. 2024b also analyzed the medium resolution data.  }  constraint on high redshift evolution of fine-structure constant with galaxy spectrum, improved the high redshift research of electromagnetism force. With the JWST data, the possible cosmic evolution of $\alpha$ and dark energy-electromagnetism coupling can be further explored, which is the main purpose of this work. 

In this paper we probe the cosmic variation of $\alpha$ using the [\ion{O}{III}] $\lambda\lambda$4959,5007 doublet emission-lines (hereafter [\ion{O}{III}]) of the very high redshift ($z>7$) galaxies. These two lines do not suffer from serious absorption and are strong enough to be reliably measured in the infrared spectrum of the high redshift objects.  
We concentrate on two sources at the redshifts of $z = 8.47$ and $7.19$ from the JWST Advanced Deep Extragalactic Survey \citep[JADES;][]{2023arXiv230602465E, 2023arXiv230602467B}. The structure of this work is as follows. In Section \ref{sec:2}, we introduce our sample, discuss the advantages of the [\ion{O}{III}] doublet to probe the possible variations of $\alpha$, and outline our method for fitting the spectroscopic data. Section \ref{sec:3} covers the calculation of the variation of $\alpha$ and the presentation of our key findings. We will wrap up and delve into the implications of our results in Section \ref{sec:4}. In Section \ref{sec:4}, we constraint the time evolution of the fine-structure constant and
Dark Energy-electromagnetism coupling. Additionally, we address potential contamination issues associated with the [\ion{O}{III}] method utilized in this study.   Throughout this work, we adopt a $\Lambda$-CDM Model with $H_0 = 67.4 {\rm \,km\,s^{-1}\,Mpc^{-1}}$, $\Omega_m = 0.3$, and $\Omega_\Lambda = 0.7$ \citep{refId0}.

\section{Sample Selection and Method}\label{sec:2}

In this section, we will first introduce our sample selection in JADES, and then we will briefly review different methods of testing the potential of varying $\alpha$ and introduce the [\ion{O}{III}] method as a probe of variation of $\alpha$. Lastly, we will describe our Markov Chain Monte Carlo (MCMC) technique to fit the galaxy [\ion{O}{III}] spectra. 

\subsection{Sample Selection}\label{sec:2.1}
In this work, we only use the JWST spectroscopic observations with F290LP-G395H filter because of its high resolution ($R = 1900 \sim$ 3600) and lower systematic uncertainty. We scan through the public JADES catalog of NIRSpec Clear-Prism Line Fluxes and NIRSpec Gratings Line Fluxes to find the sources with redshift($z$) over 7.1, and set a Signal-to-Noise Ratio (SNR) test. Letting the SNR for the [\ion{O}{III}]$\lambda$5007 line intensity be at least 8:1 which means the area under the 5007 \AA \ line is measured to an accuracy of $\pm 12.5 \%$, we finally select the two ELGs and their NIRSpec ID are 00008013 and 10013905, respectively. 

\subsection{Probe the $\alpha$ variation with the [\ion{O}{III}] doublet}\label{sec:2.2}
In the early exploration of the time variation of $\alpha$, astronomers initially adopted the emission lines of nearby quasars \citep{Savedoff1956, bahcall1965interaction} to establish an upper limit. However, this approach was soon abandoned because of the broad emission lines can be influenced by the central supermassive black holes and researchers decided to focus on studying quasar absorption lines to understand the time dependence of $\alpha$ \citep{bahcall1967analysis}.  Subsequently, the alkali-doublet (AD) method was developed to analyze absorption lines from gas clouds utilizing luminous quasars as reference points. While effective, the AD method does not fully exploit the potential information in quasar spectra, leading to the development of the many-multiplet (MM) method. This method leverages multiple atomic transitions from various multiplets and ionization stages to enhance sensitivity in probing $\alpha$'s time dependence \citep{Webb_1999, Webb_2001}. However, the MM method introduces uncertainties due to the complex many-body atomic theory involved \citep{bahcall2004does}.

Thus, we take the [\ion{O}{III}] emission lines in ELGs as a probe of varing $\alpha$ in the early universe. There are some advantages of using [\ion{O}{III}] to probe the varying $\alpha$. Firstly, the doublet lines have a significantly wider wavelength separation of around 50 \AA , which is approximately ten times larger than the separation seen in most fine-structure doublet lines. Besides, the [\ion{O}{III}] doublet lines are much narrower and cleaner in emission-line galaxies (ELGs), and sometimes the strongest in spectra, so it is easy for us to get high SNR data of the [\ion{O}{III}]. The whole procedure of detection is also simple and straightforward without making assumptions in MM method\citep{bahcall2004does}. High redshift observations ($z \gtrsim 7$) pose challenges for traditional methods like MM and AD due to neutral hydrogen absorption and limited heavy elements in the early universe \citep{1998ApJ...501...15M, Robertson2010, Wise_2019}. While ELGs offer a promising avenue for studying $\alpha$ variation, the scarcity of high-resolution ELG spectra hampers their widespread use.

Considering non-relativistic approximation, the observed separation of wavelength ($\Delta \lambda = \lambda_1 - \lambda_2$) of the doublet is proportional to the fine structure constant: $\Delta \lambda / \bar{\lambda} \propto \alpha^2$.   Here $\bar{\lambda}$ is the average of the wavelengths. Then the anomaly of the fine structure constant $\Delta \alpha / \alpha$ is calculated by 
\begin{eqnarray}\label{eq1}
\frac{\Delta \alpha}{\alpha} = \sqrt{\frac{\Delta \lambda(z) / \bar{\lambda}(z)}{\Delta \lambda(0) / \bar{\lambda}(0)}} - 1,
\end{eqnarray}
where $z$ stand for the source redshift. The denominator, the present-day value of $\Delta \lambda / \bar{\lambda}$, is \citep{bahcall2004does, peck1972dispersion, moorwood1980infrared, pettersson1982spectrum} 
\begin{eqnarray}
R(0) = \frac{\Delta \lambda(0)}{\bar{\lambda}(0)} = 4.80967 \times 10^{-3}[1 \pm 0.00001].
\end{eqnarray}

\subsection{MCMC method to fit the spectrum}
In order to fit in the spectroscopic data, we construct a model to describe the shape of [\ion{O}{III}] lines and the continuum component in the spectra, which reads 
\begin{eqnarray}
F(x|a, b, \lambda_2, \Delta \lambda, n_1, n_2, s_1, s_2) = - 10^a x + 10^b \nonumber \\
      + \frac{10^{n_1 - s_1}}{\sqrt{2 \pi}} e^{\frac{-\left(x - \left(\lambda_2 - \Delta \lambda\right)\right)^2}{\left(2 \times 10^{s_{1}}\right)^2}}  
      +\frac{10^{n_2 - s_2}}{\sqrt{2 \pi}} e^{\frac{-\left(x - \lambda_2\right)^2}{\left(2 \times 10^{s_{2}}\right)^2}},
      \label{eq:3}
\end{eqnarray}
where $a$ is the log-slope of the continuum, $b$ is the log-intercept of the continuum, $\lambda_2$ is the wavelength of the [\ion{O}{III}]$\lambda$5007 line, $\Delta\lambda$ is the wavelength separation between the two [\ion{O}{III}]$\lambda$4959 and [\ion{O}{III}]$\lambda$5007 fine structure lines, $n_1$ and $s_1$ are the two parameters amplitude and 1-$\sigma$ width of the [\ion{O}{III}]$\lambda$4959 Gaussian emission line profile, $n_2$ and $s_2$ are the 
same component of the [\ion{O}{III}]$\lambda$5007 emission line profile. The slope of the spectral continuum emission is negative, which is because of the strong ultraviolet (UV) radiation continuum emission. This UV radiation is a result of their vigorous star formation activities \citep{topping2024uv}. 
One linear function is to fit the background and two Gaussian-like functions are designed for fitting the [\ion{O}{III}] line shape. There are eight free parameters in this model and we define an objective function
\begin{eqnarray}
\chi^2 = \sum_{i = 1}^N \frac{(y_i - F_i)^2}{\sigma_y^2}, 
\end{eqnarray}
and a log-likelihood used to estimate the posterior probabilities
\begin{equation}
-2\ln{\mathcal{L}} = \chi^2. 
\end{equation}
Our goal is to maximize this $\mathcal{L}$ and we deploy the MCMC technique to estimate all the eight parameters with the help of {\tt emcee} package\citep{2013PASP..125..306F}. The $priori$ range of the eight free parameters are listed in Tabel~\ref{tab:prior}.

\begin{table}
\bc
\begin{minipage}[]{100mm}
\caption[]{The $priori$ range of the eight free parameters.}\label{tab:prior}\end{minipage}
\setlength{\tabcolsep}{2.5pt}
\small
 \begin{tabular}{ccccccccccccc}
  \hline\noalign{\smallskip}
Parameter &  type & range \\
  \hline\noalign{\smallskip}
$a$ & Uniform & ($-23.0$, $0$) \\
$b$ & Uniform & ($-29$, $0$) \\
$\frac{\lambda_{5007}}{(1+z)}$ & Uniform & ($4996.84$, $5016.84$) \\
$\frac{\Delta\lambda}{(1+z)}$ & Uniform & ($38$, $58$) \\
$n_1$ & Uniform & ($-24.5$, $-21$) \\
$n_2$ & Uniform & ($-23$, $-20$) \\
$s_1$ & Uniform & ($-4$, $-1.5$) \\
$s_2$ & Uniform & ($-4$, $-1.5$) \\
  \noalign{\smallskip}\hline
\end{tabular}
\ec
\end{table}

After 50000 times run, we put the posterior parameters into the $priori$ and run another 50000 times. Cornerplots of the full set of posterior distributions are shown below in the Appendix \ref{A} and the final parameter estimation(Table \ref{Tab:1}) are also laid down.

\begin{table}
\bc
\begin{minipage}[]{100mm}
\caption[]{Final parameter estimation of the two [\ion{O}{III}] emission line galaxies with 1-$\sigma$ of the posterior as statistical uncertainty}\label{Tab:1}\end{minipage}
\setlength{\tabcolsep}{2.5pt}
\small
 \begin{tabular}{ccccccccccccc}
  \hline\noalign{\smallskip}
Parameters& NIRSpec 10013905& NIRSpec 00008013 \\
  \hline\noalign{\smallskip}
$\lambda_2(\mu m)$&$4.11^{+0.0000267}_{-0.0000266}$ &$4.74^{+0.0000668}_{-0.0000663}$\\
$\Delta \lambda(\mu m)$&$0.0393^{+0.0000662}_{-0.0000649}$&$0.0453^{+0.000162}_{-0.000164}$\\
  \noalign{\smallskip}\hline
\end{tabular}
\ec
\end{table}

\section{Results}\label{sec:3}

\begin{figure} 
   \centering
   \includegraphics[width=12.0cm, angle=0]{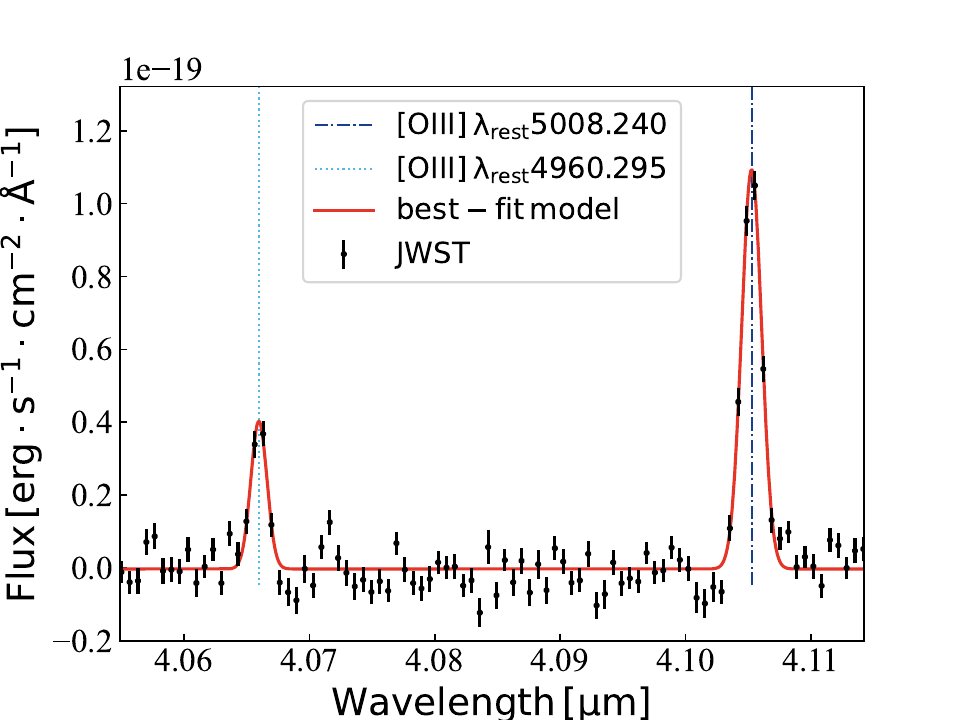}
   \caption{The best-fit model and JWST-NIRSpec data of NIRSpec 10013905 [\ion{O}{III}] emission lines. The black points with the error bars are the observed JWST spectrum data. The red line is the best-fit model of the [\ion{O}{III}] doublet emission lines. The dotted line in cyan and dot-dashed line in blue are best-fit lines' wavelength of [\ion{O}{III}]$\lambda$4959 and [\ion{O}{III}]$\lambda$5007 lines.  } 
   \label{fig:1}
   \end{figure}

\begin{figure}
    \centering
    \includegraphics[width=12.0cm, angle=0]{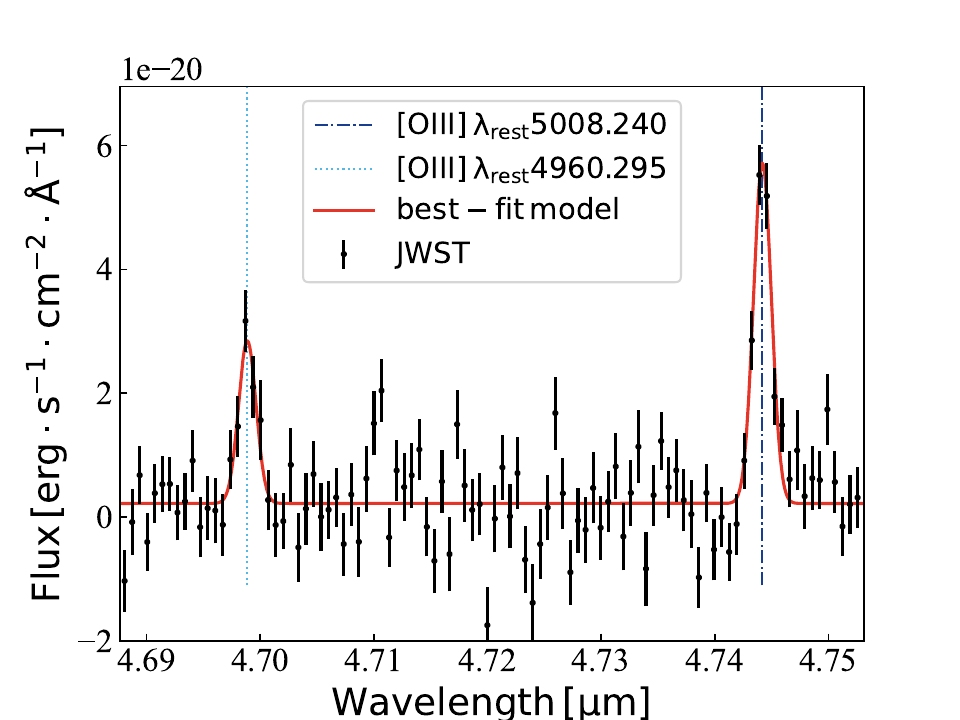}
    \caption{Same as Figure~\ref{fig:1}, but for the source NIRSpec 00008013.}
    \label{fig:2}
\end{figure}

In this section, we will present the final results of the best-fit spectrum profiles, and calculate the $\Delta \alpha / \alpha$. Substituting the fitted results in Table \ref{Tab:1} to the equation (\ref{eq1}), we work out the result of $\frac{\Delta \alpha}{\alpha}$. And the best-fit line profiles are displayed in Figure \ref{fig:1} and Figure \ref{fig:2} for our two sources. 

However, we also need to take into account the systematic uncertainties. The systematic uncertainty due to the spectral resolution is $\delta \lambda = \frac{0.00066498\, (\mu m)}{50} =1.33\times10^{-5}\, (\mu m)$ in this wavelength range, where the factor $\frac{1}{50}$ is the JWST spectral calibration error \citep{2022A&A...661A..80J}. Here, the systematic error is attributed to the imaging calibration of the NIRSpec filter and grating wheels. \footnote{ The relative wavelength calibration, which takes into account the effects of dispersion, is a complex process and has not yet yielded robust results. Consequently, we have estimated the systematic uncertainty associated with the imaging calibration of the NIRSpec filter and grating wheels.} 
Because the systematic errors of $\Delta \lambda$ and $\lambda_2$ are from the same resolution limit, these two parameters have the same systematic uncertainty equal to $\delta \lambda$.
Throwing these factors into the propagation equation of uncertainty
\begin{eqnarray}
 \sigma^2_{\frac{\Delta \alpha}{\alpha}} &=& \frac{\Delta \lambda}{4 R^3(0)(- \Delta \lambda + 2 \lambda_2)} \sigma^2_{R(0)} \nonumber \\
 &+& \frac{\Delta \lambda}{R(0)(- \Delta \lambda + 2 \lambda_2)^3} \sigma^2_{\lambda_2} \nonumber\\
 &-& \frac{\lambda_2^2}{\Delta \lambda(\Delta \lambda - 2 \lambda_2)^3 R(0)} \sigma^2_{\Delta \lambda}, 
\end{eqnarray}
we have $0.44^{+8.4(stat)+1.7(sys)}_{-8.3(stat)-1.7(sys)} \times 10^{-4}$ and $-10.0^{+18(stat)+1.5(sys)}_{-18(stat)-1.5(sys)} \times 10^{-4}$ for these two sources NIRSpec 10013905 and NIRSpec 00008013, respectively. The larger uncertainty is the systematic one. The smaller one is 1-$\sigma$ statistical error from the fitting. It is apparent that 
the systematic uncertainty plays the dominant role in 
constraining the $\alpha$ variation. Thus, higher resolution observations are needed for further investigation. Or, we have to identify more distinct emission lines in the spectra of the objects in the early universe. 

\section{Conclusions and discussions}\label{sec:4}

To conclude, we utilized the farthest observations of the [\ion{O}{III}] emission lines to constrain the variability of $\frac{\Delta \alpha}{\alpha}$. This choice was primarily due to the strength of these two lines, making them detectable in the high redshift Universe, and the wide enough separation between the lines for the instrument to capture. These emission lines act as a natural probe for researchers to explore the possibility of a changing $\alpha$. Our findings suggest that $\alpha$ has remained nearly constant since the early Universe.

\subsection{Constraint on the time evolution of the fine structure constant}

Combining the results before, the final sample of $\frac{\Delta \alpha}{\alpha}$ spans the redshift range of $0.2 < z < 8.5$. The data are illustrated in Figure \ref{fig:4}. The cited data at lower redshifts $0.2<z<7.1$ are from Refs. \citep{10.1111/j.1365-2966.2012.20852.x,10.1093/mnras/stv2148,martins2017stability,doi:10.1126/sciadv.aay9672}.

We use these data to estimate the relative time variance of the fine structure constant ($\frac{1}{\alpha} \frac{d \alpha}{dt}$) via the procedure developed by \cite{bahcall2004does}. 
We fit the linear function of time
\begin{eqnarray}\label{eq7}
 \frac{\Delta \lambda(t(z)) / \bar{\lambda}(t(z))}{\Delta \lambda(0) / \bar{\lambda}(0)} = 1 + S H_0 t. 
\end{eqnarray}
Here we use the well-known expression
\begin{eqnarray}
 t = \int_{0}^{z}\frac{(1 + z^{'})^{-1}dz^{'}}{\sqrt{(1 + z^{'})(1 + \Omega_m z^{'}) - z^{'}(2 + z^{'})\Omega_\Lambda}}
\end{eqnarray}
to convert $z$ to $t$ \citep{mukhanov2005physical}. The slope $S$ in eq.(\ref{eq7}) is
\begin{eqnarray}
 S = \frac{1}{H_0 \alpha^2}\left(\frac{d\alpha^2}{dt}\right) = \frac{2}{H_0 \alpha}\frac{d\alpha}{dt}. 
\end{eqnarray}
We determine the parameters with MCMC method once again. Our result is
\[\frac{1}{\alpha} \frac{d \alpha}{dt} = 0.30^{+4.5}_{-4.5}\times 10^{-17}~{\rm yr^{-1}} .
\]
The statistical uncertainty is 1-$\sigma$(the bold ones are 2-$\sigma$) and the postierior distribution is shown underneath in appendix \ref{A}. It again indicates an time-independent $\alpha$ in our universe. 

\subsection{Constraint on Dark Energy coupling with electromagnetism}

The coupling between dark energy and electromagnetism could potentially lead to observable phenomena, such as the temporal evolution of the fine structure constant \citep{Avelino:2006gc,Thompson:2013xda,Calabrese:2013lga,Chen_2019}. The interaction between the dark energy scalar field ($\phi$) and electromagnetism ($F$) can be described by a gauge kinetic function $B_{F(\phi)}$ \citep{Calabrese:2013lga}:
\begin{eqnarray}
\mathcal{L}_{\phi F} = -\frac{1}{4} B_F(\phi) F_{\mu \nu} F^{\mu \nu}.
\end{eqnarray}

This gauge kinetic function is linearly dependent on the scalar field \citep{Avelino:2006gc,Nunes:2003ff,Vielzeuf:2013aja}:
\begin{eqnarray}
B_F(\phi) = 1 - \zeta \sqrt{8 \pi G} (\phi - \phi_0), \label{eq:gauge}
\end{eqnarray}
where $\zeta$ represents the coupling strength between the scalar dark energy field and the electromagnetic field.

Consequently, the variation of the fine structure constant $\alpha$ due to the dark energy-electromagnetism coupling is expressed as \citep{Calabrese:2013lga}:
\begin{eqnarray}
\frac{\Delta \alpha}{\alpha} \equiv \frac{\alpha - \alpha_0}{\alpha_0} = \zeta \sqrt{8 \pi G} (\phi - \phi_0). \label{eq:gauge2alpha}
\end{eqnarray}

The temporal evolution of the fine structure constant $\alpha$, attributed to the coupling between dark energy and electromagnetism, can be reformulated as follows \citep{Calabrese:2013lga}:
\begin{eqnarray}
\frac{\Delta \alpha}{\alpha}(z) = \zeta \int_0^z \sqrt{3 \Omega_\phi(z)[1+w(z)]} \frac{d z^{\prime}}{1+z^{\prime}}. \label{eq:alpha_vary}
\end{eqnarray}

There is no simple analytical solution to the equation of state of the dark energy scalar field mentioned above, and most of the previous work (e.g \cite{Avelino:2006gc,Nunes:2003ff}) has used higher-order polynomials for fitting or modeling, and some works (e.g \cite{Vielzeuf:2013aja,Calabrese:2013lga}) has assumed a parameterized Chevallier-Polarski-Linder (CPL) dark energy equation of state, and then performed a parameterized fitting of the change of the fine structure constant.
To constrain the coupling strength $\zeta$ between dark energy and electromagnetism, we adopt the Chevallier-Polarski-Linder (CPL) parametrization of the dark energy equation of state, which features two adjustable parameters, $w_0$ and $w_a$ \citep{Chevallier:2000qy,Linder:2002et}:
\begin{eqnarray}
w(z) = w_0 + w_a \frac{z}{1+z}. \label{eq:DE-EOS}
\end{eqnarray}

The fractional density of dark energy in the universe is articulated by:
\begin{eqnarray}
\Omega_{\phi}(z) = \Omega_{\phi0} \left(1+z\right)^{3\left(1+w_0 +w_a\right)} e^{-3w_a\frac{z}{1+z}}. \label{eq:DE-rho}
\end{eqnarray}

\begin{figure}
    \centering
    \includegraphics[width=14.0cm, angle=0]{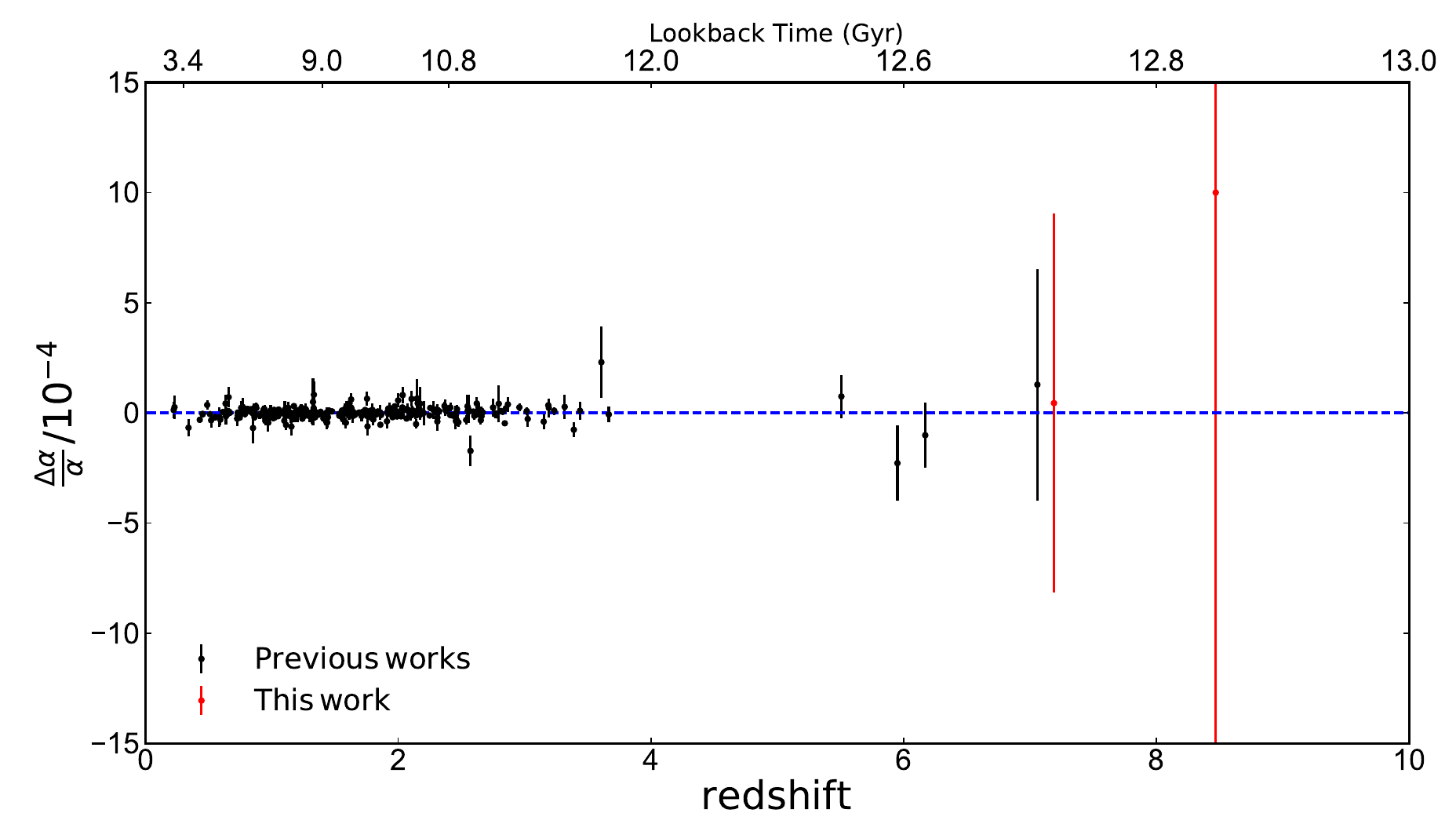}
    \caption{Direct measurements of $\frac{\Delta \alpha}{\alpha}$ in different cosmic epochs. The red data points are from best-fit [\ion{O}{III}] results of the two JWST emission line galaxies NIRSpec 10013905 and NIRSpec 00008013 in this work. The others in black are from different references with MM methods \citep{10.1111/j.1365-2966.2012.20852.x,10.1093/mnras/stv2148,martins2017stability,doi:10.1126/sciadv.aay9672}. }
    \label{fig:3}
\end{figure}

The black data points in Figure \ref{fig:3} represent previous findings, while the current constraints from the James Webb Space Telescope (JWST) on $\frac{\Delta \alpha}{\alpha}(z)$ are depicted as red error bars.

We have integrated the latest JWST [\ion{O}{III}] measurement of $\Delta\alpha/\alpha$ with prior measurements cited in \cite{10.1111/j.1365-2966.2012.20852.x,10.1093/mnras/stv2148,martins2017stability,doi:10.1126/sciadv.aay9672} to establish constraints on the dark energy-electromagnetism coupling as defined by Equations (\ref{eq:gauge}) to (\ref{eq:DE-rho}). During the fitting process, we assigned Gaussian priors to the parameters $w_0$ and $w_a$: $\mathcal{G} (-0.957\pm0.08)$ for $w_0$ and $\mathcal{G} (-0.29\pm0.3)$ for $w_a$, reflecting observational constraints from Planck18+SNe+BAO \cite{refId0}. For the dark energy-electromagnetism coupling strength $\zeta$, we imposed a log-flat prior with an upper limit of $\log_{10}\zeta=1.0$ and a lower limit of $\log_{10}\zeta\approx -38.2$. This lower boundary corresponds to the dimensionless gravitational fine-structure constant $\alpha_g = \frac{G_N m^2_p}{\hbar c} \approx 6\times 10^{-39}$ as derived in \cite{MOSS_2010}.

\begin{figure}
    \centering
    \includegraphics[width=7.0cm, angle=0]{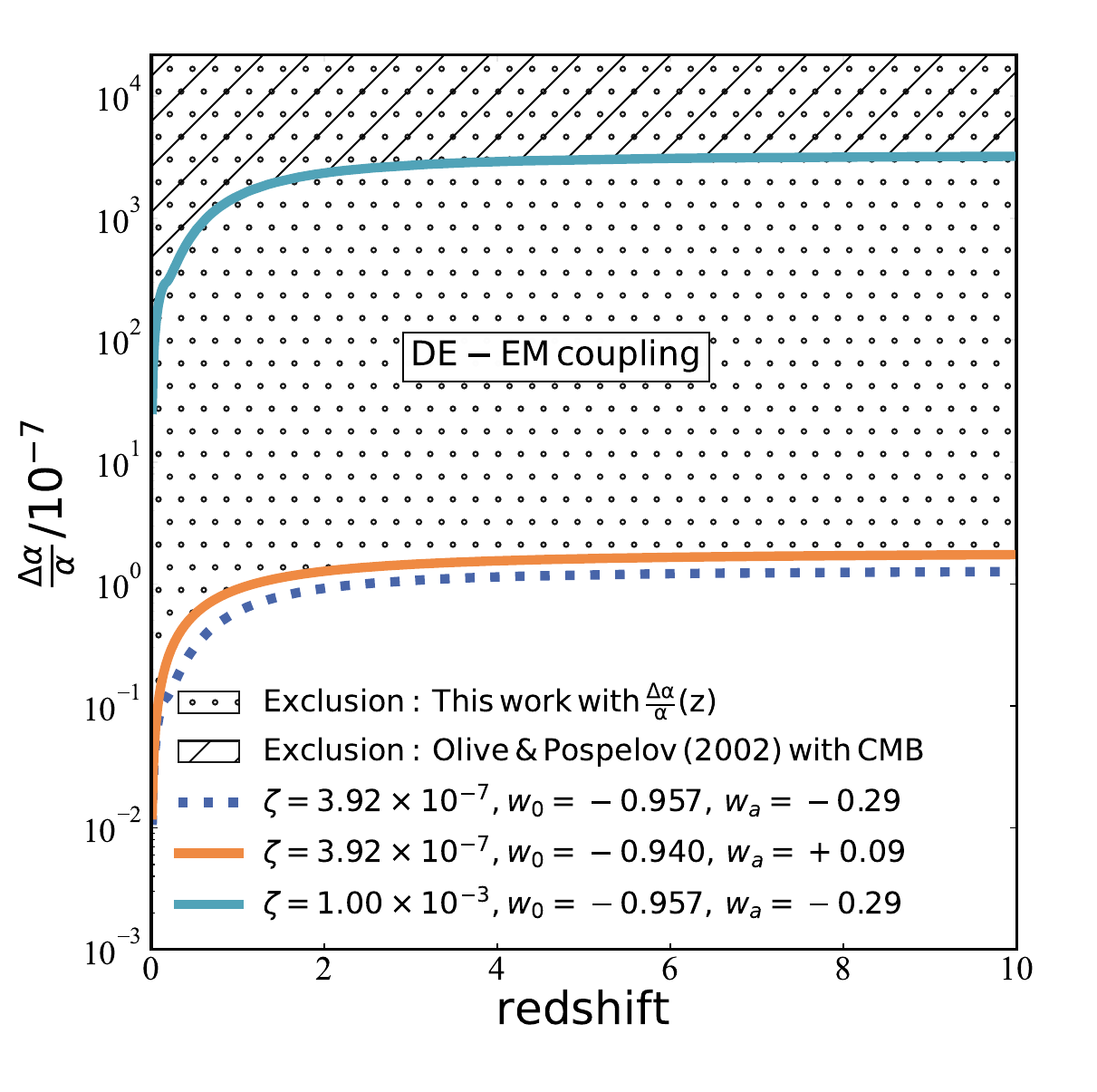}
    \includegraphics[width=7.0cm, angle=0]{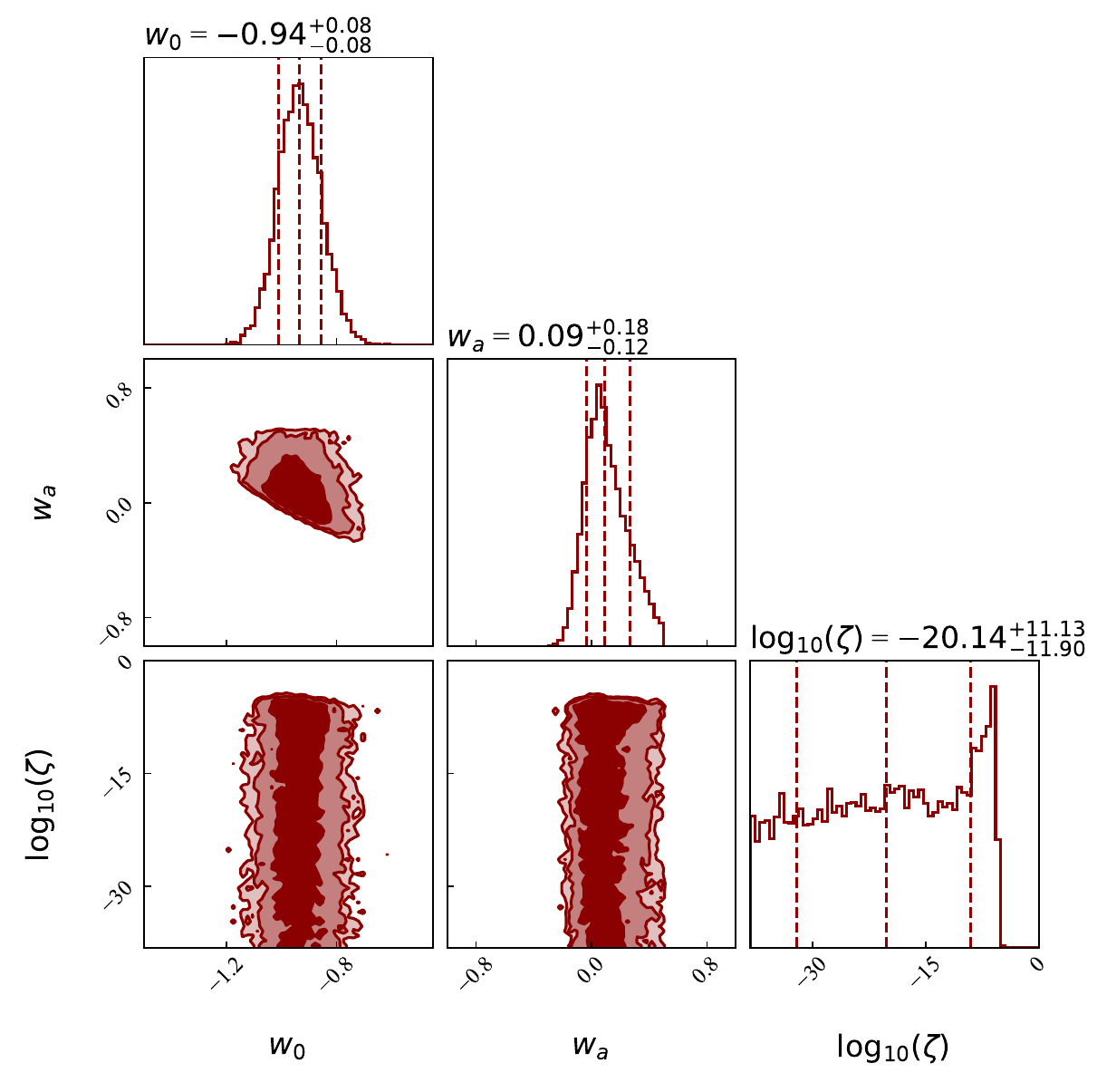}
    \caption{Left: The constraints on fine-structure constant evolution induced by dark energy coupling with electromagnetism.  The lines in different styles and colors are constraints corresponds to different parameters. The area filled by the slash is the previous exclusion of $\zeta<10^{-3}$ with CMB by \cite{Olive:2001vz}. The area filled by the dot is exclusion of this work. The parameter $w_0$, $w_a$ are free parameters of dark energy equation of state in Equation~(\ref{eq:DE-EOS}). The parameter $\zeta$ is dark energy-electromagnetism coupling strength in Equation~(\ref{eq:alpha_vary}).  Right: The posterior distribution of parameters of the dark energy-electromagnetism coupling model fitting to the combined $\frac{\Delta\alpha}{\alpha}(z)$ measurements.  }
    \label{fig:4}
\end{figure}

The colored lines in Figure \ref{fig:4} illustrate the evolution of the fine-structure constant according to the dark energy-electromagnetism (DE-EM) coupling model. The lines with various line styles represent the variation of $\frac{\Delta \alpha}{\alpha}(z)$ for different strengths of DE-EM coupling. The current collective data of $\frac{\Delta\alpha}{\alpha}(z)$ has led to a stringent constraint on the dark energy-electromagnetism coupling strength $\zeta$:
\[
\zeta \leq 3.92 \times 10^{-7} \; (\text{at \; 95\% \; Confidence \; Level}). \label{eq:DE-EM+constraint}
\]
This constraint is contingent upon the utilization of the CPL dark energy equation of state. Additionally, we have examined the constraints for the dark-energy-electromagnetic coupling model under various types of priors, such as flat priors. The outcomes indicate that the results are consistent with those aforementioned, with the constraints derived from different priors exhibiting similar magnitudes (i.e. orange solid line v.s. blue dotted line in Figure.~\ref{fig:4}). The current 95\% limit is about three orders of magnitude stronger than the results of other previous work (e.g., $\zeta<10^{-3}$ obtained by  \cite{Olive:2001vz} with with cosmic microwave background (CMB)).

\subsection{Excluding the potential contamination of the [\ion{O}{III}] method}

Although the [\ion{O}{III}] emission lines have advantages in testing the variation of $\alpha$ in the high redshift Universe (i.e., see Section~\ref{sec:2.2}), it may be not precise because of the pollution from the galaxies with active galactic nucleis (AGN, or quasar). The quasar emission lines tend to be broad, which is not ideal for precise measurement of radial velocity. While the [\ion{O}{III}] lines in quasars are not as broad as lines like [\ion{C}{IV}] and [\ion{Mg}{II}], they are still broader than those typically seen in galaxies. This broadening can lead to contamination of the H$\beta$ lines, making it challenging to accurately determine the center of the lines \citep{bahcall2004does, 2024ApJ...968..120J}. We cross-matched the current catalogs \citep{2023arXiv230801230M,2023arXiv231118731S} of AGNs including the JADES data, there is one 
potential AGN in our sample. The source NIRSpec-10013905 was identified as an $z\sim7.2$ AGN candidate by \cite{2023arXiv231118731S} because of its [\ion{He}{II}]$\lambda 4686$ emission line. This optical He emission line maybe from the ionized He by the ionising radiation due to the central AGN. However, we find no strong evidence of the width of the [\ion{O}{III}] doublet emission lines influenced by the AGN under the current uncertainty. Thus, that is not important to the current result.

\normalem
\begin{acknowledgements}
This work is supported by the Natural Science Foundation of China (No. 11921003, No. 12233011, No. 12373002).
\end{acknowledgements}

\appendix
\section{Cornerplots of the MCMC fitting}\label{A}

We produce the posterior distributions with Equation~(\ref{eq:3}). The eight free parameters are $a, b, \lambda_2, \Delta \lambda, n_1, n_2, s_1, s_2$. Clearly, the $\lambda_2, \Delta \lambda, n_1, n_2, s_1, s_2$ are almost normally distributed. The plots only included the statistic uncertainties from the fitting. The analysis of systematic error is discussed in Section~\ref{sec:3}. 

\begin{figure*}[ht]
    \centering
    \includegraphics[width=12.0cm, angle=0]{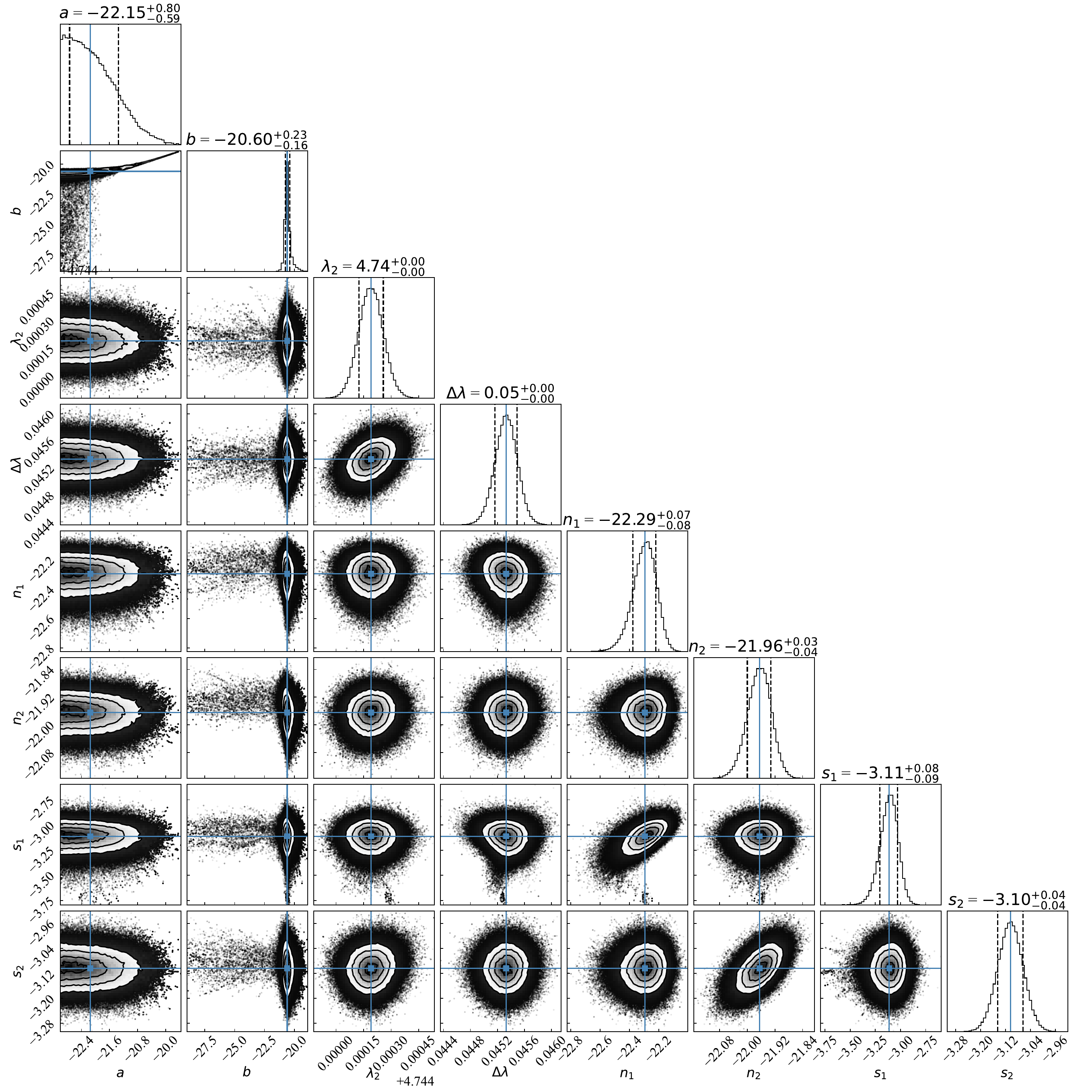}
    \caption{Posterior distribution of the parameters of NIRSpec 00008013. The diagonal figures are the posterior distributions of each parameters and the rest are joint distributions of one another. The blue solid lines are the 50\% quantiles, and the dashed lines are for the $1\sigma$ range. }
    \label{fig:A1}
\end{figure*}

\begin{figure*}[ht]
    \centering
    \includegraphics[width=12.0cm, angle=0]{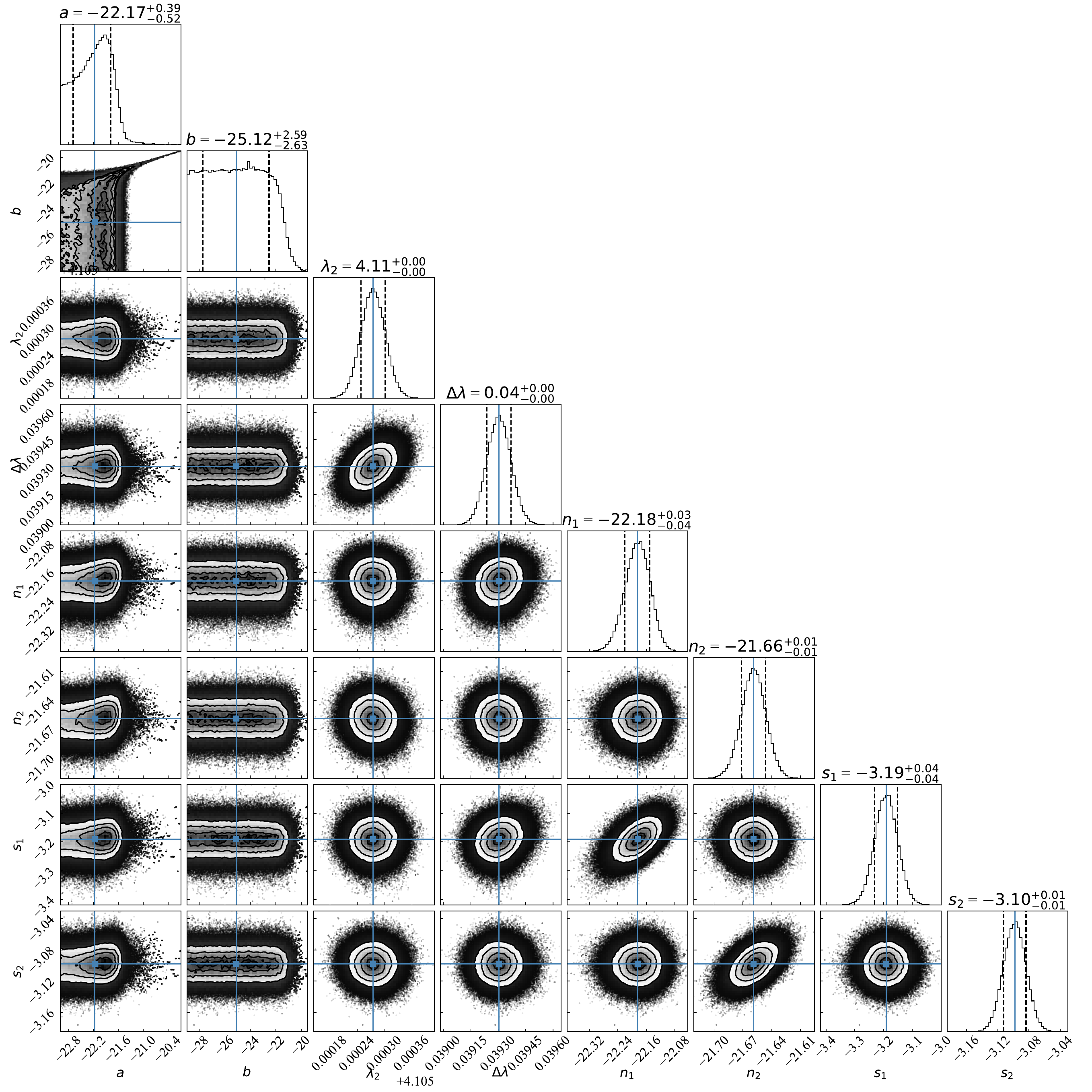}
    \caption{Posterior distribution of the eight parameters of NIRSpec 10013905.}
    \label{fig:A2}
\end{figure*}

\begin{figure}[ht]
    \centering
    \includegraphics[width=12.0cm, angle=0]{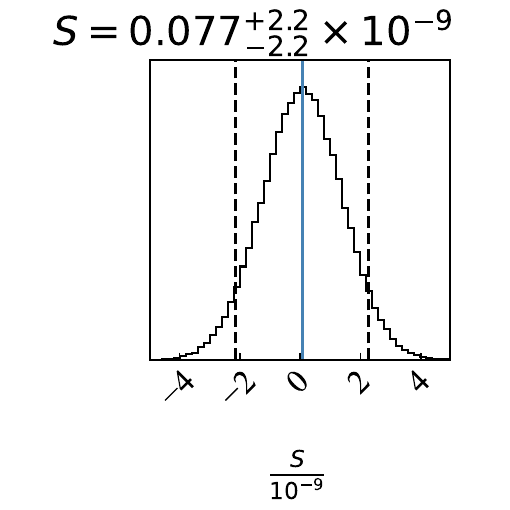}
    \caption{Posterior distribution of S. The blue solid line is the 50\% quantile, and the dashed lines are for the $1\sigma$ range.}
    \label{fig:A3}
\end{figure}

\clearpage
  
\bibliographystyle{raa}
\bibliography{bibtex}

\end{document}